\documentclass[prb,twocolumn,showpacs,preprintnumbers]{revtex4}%
\usepackage{graphicx}
\usepackage{dcolumn}
\usepackage{bm}
\usepackage{amsmath}
\usepackage{amsfonts}
\usepackage{amssymb}
\providecommand{\U}[1]{\protect\rule{.1in}{.1in}}
\begin{document}
\title{Experimental evidence for nematic order of cuprates in relation to lattice structure}
\author{David J. Singh}
\affiliation{Materials Science and Technology Division, Oak Ridge National Laboratory, Oak
Ridge, Tennessee 37831-6114}
\author{I. I. Mazin}
\affiliation{Code 6690, Naval Research Laboratory, Washington, DC 20375}
\date{\today}

\begin{abstract}
Experiments that have been interpreted as providing evidence that the
pseudogap phase in cuprates is an electronic nematic are discussed from the
point of view of lattice structure.
We conclude that existing experiments are not sufficient to prove a
nematic origin of the pseudogap in underdoped cuprates.
\end{abstract}

\maketitle

\section{Introduction}

Understanding the nature of the pseudogap
\cite{warren,alloul,ding,tallon,norman} and its relationship to
superconductivity has emerged as a central problem in high $T_{c}$ cuprate
superconductivity. While there has been considerable progress in elucidating
its momentum and doping dependence, its microscopic origin and relationship to
superconductivity remain controversial. Some widely discussed experimentally
supported possibilities are (1) that it is associated with an order with a
quantum critical point below the superconducting dome that may play a key role
in superconductivity and normal state transport. \cite{varma,li} (2) that it
represents Cooper pairing without phase coherence, i.e. a strong coupling
precursor of superconductivity, \cite{kanigel} and in contrast (3) that it is
a phase associated with cuprates that competes with superconductivity.
\cite{yoshida,kondo} There have also been recent discussions, also based on
experimental results, of the pseudogap phase as an electronic nematic,
specifically associating the pseudogap temperature $T^{*}$ with an onset of
electronic nematicity. \cite{fradkin,kivelson,vojta,daou,hackl} The purpose of
this paper is to discuss the experimental evidence for this. We argue that,
contrary to recent suggestions, \cite{fradkin} the evidence for an association
of $T^{*}$ with an onset of nematic order is actually weak.

We emphasize that we are not discussing the so-called stripes in the strongly
underdoped 214 cuprates. Theoretically, these are a dynamic phase separation
that minimizes the Coulomb energy, and they occur on a larger length scale
that the microscopic nematicity that is thought to occur in Sr$_{3}$Ru$_{2}%
$O$_{7}$ and above $T_{AFM}$ in pnictides.

Specifically, we analyze the statement that at the temperature $T^{*}$ where a
pseudogap in the excitation spectrum develops the symmetry the CuO$_{2}$
plane(s) breaks from a four-fold (tetragonal) to at most a two-fold. The
experiments that seem to support this idea include transport measurements on
YBa$_{2}$Cu$_{3}$O$_{7-\delta}$, \cite{daou} neutron scattering experiments on
YBa$_{2}$Cu$_{3}$O$_{6.45}$ \cite{hinkov} and especially scanning probe
measurements on several materials. \cite{kohsaka-2,kohsaka-3}.

\begin{figure}[ptb]
\includegraphics[width=.32\columnwidth]{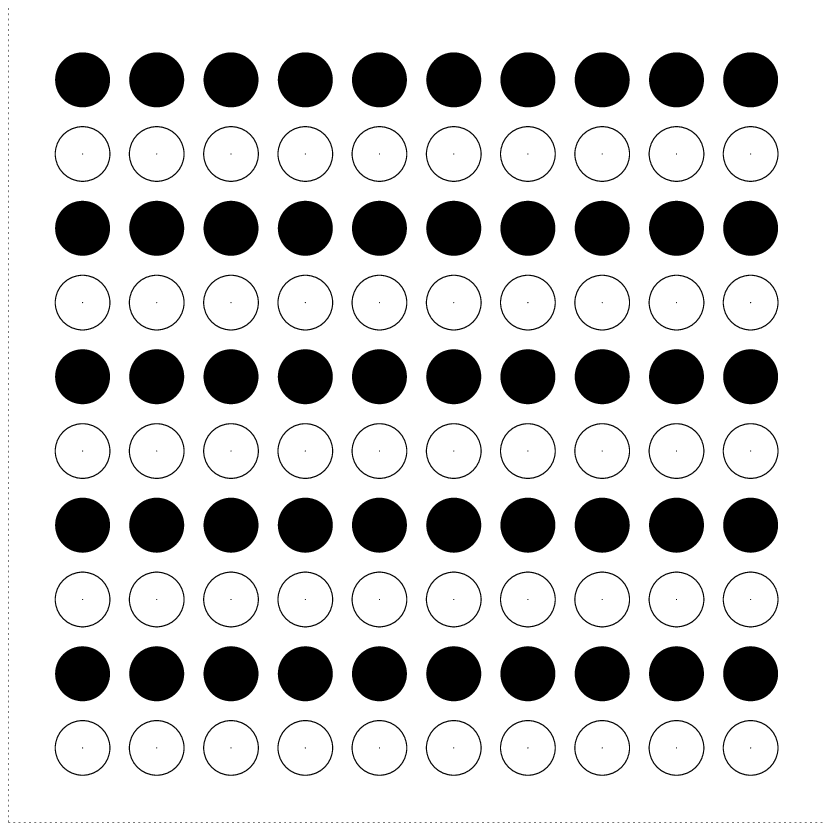}
\includegraphics[width=.32\columnwidth]{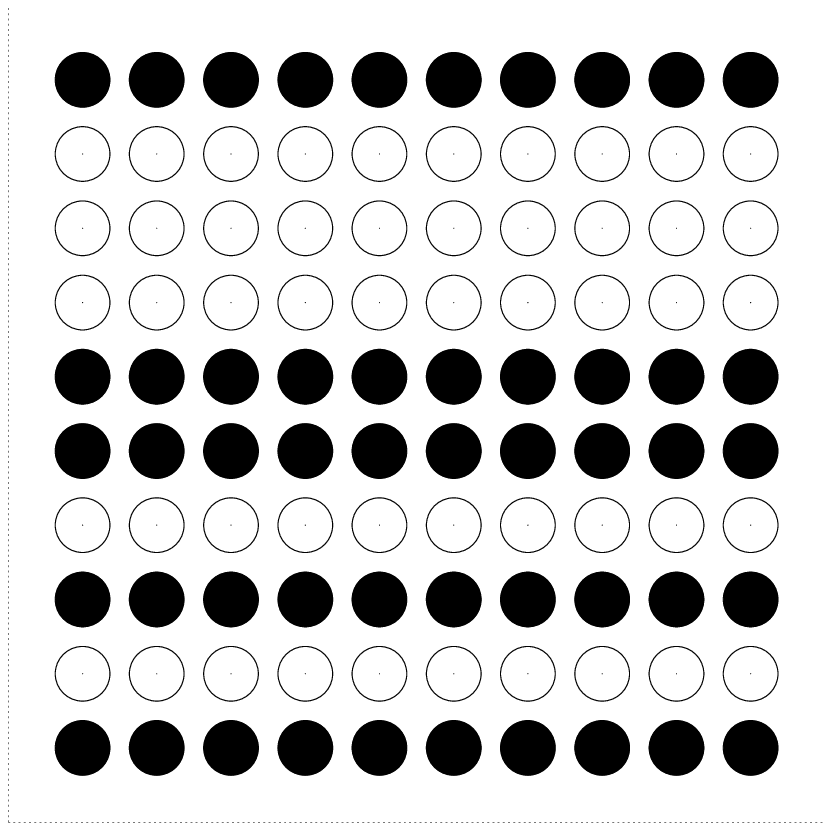}
\includegraphics[width=.32\columnwidth]{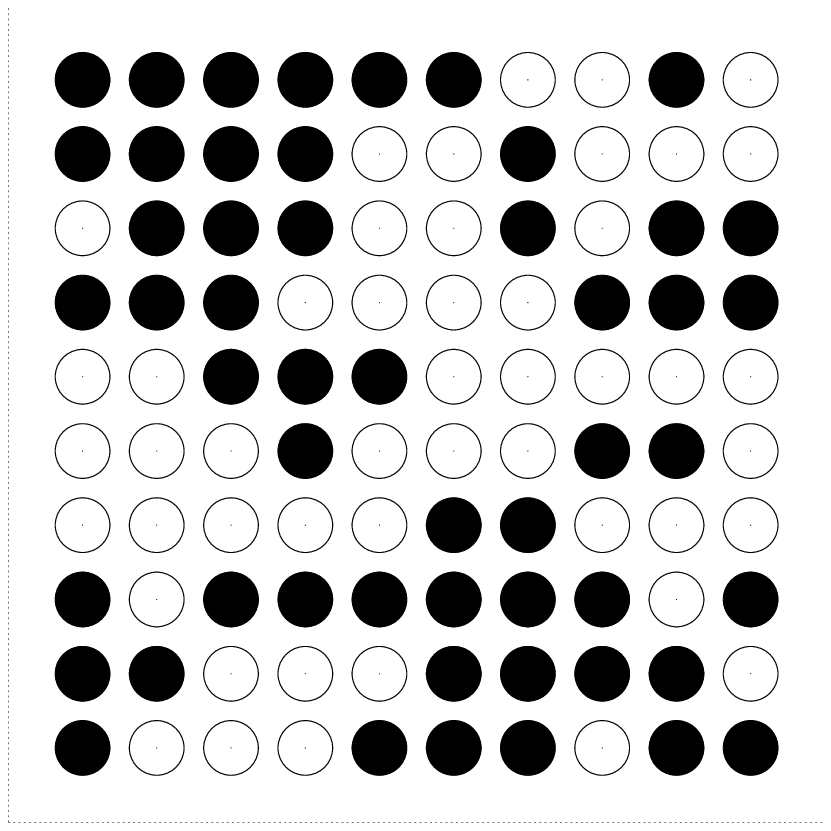}
\caption{Schematic depiction of
different types of ordering in crystalline media: (a) Symmetry-breaking long
range order; (b) Symmetry-breaking texture (long range order in one direction
only); (c) nematic order. In the last case, there is no long-range ordering in
any direction but the number of the like bonds along $x$ is different from the
number of unlike bonds, while along $y$ there are the same number of like and
unlike bonds.}%
\label{nematic}%
\end{figure}

We begin with definitions, which are important in this area because research
on nematicity in strongly correlated systems is quickly developing field and
perhaps as a result the term has become broadly applied in some cases to
phenomena that are rather common and do not fit under the standard
definitions. The formal definition of a nematic phase is found in the theory
of liquid crystals. \cite{lq} There a mesomorphic (lacking long range order)
liquid crystal is called $nematic$ if the constituent anisotropic molecules
have one preferential orientation axis. At the same time, the centers of
molecules are disordered. In this sense, a nematic solid state phase would be
a glass or a disordered alloy where short-range correlations in one direction
are different from the other. This is illustrated in Fig. \ref{nematic}. The
left panel shows an ordered orthorhombic crystal; note that the lattice
parameters $a=b$, but the crystal symmetry is orthorhombic and a long range
order is present. Such as state is not a nematic. Note also that the two
species need not be different ions, they can just as well be spins on
different sites. The middle panel shows a textured crystal where there is a
long range order in one direction, but not in the other. Strictly speaking,
this is not a nematic phase either. On the other hand, the right panel is an
example of a canonical nematic phase: there is no long range order at all, yet
the number of the like bonds along $a$ is 70\% larger than the number of the
unlike bonds, while along $b$ there is no such correlation. Such a crystal may
show substantial anisotropy in $e.g.$ transport properties. In fact, this is
the signature of nematicity -- a breaking of rotational symmetry without
translational symmetry breaking.

Nematicity in the electronic subsystem may or may not trigger a significant
difference between $a$ and $b$ depending on the relative energy scales
associated with the nematic order and the lattice (i.e. elastic constants).
For instance, a considerable anisotropy in transport is observed in the high
field nematic phase of Sr$_{3}$Ru$_{2}$O$_{7}$ without any detectable $a/b$
differentiation. \cite{borzi} On the other hand, there is a strong structural
signature at the onset of the very likely nematic phase that precedes the spin
density wave transition as as LaFeAsO is cooled. \cite{cruz,mazin-fe} However,
note that the spin density wave phase itself has long range translational
symmetry breaking, and while it lowers the rotational (point group) symmetry
from tetragonal to orthorhombic, it is not a nematic.

To summarize, the concept of electronic nematicity can and should be applied
to crystalline solids to describe a situation where electronic or magnetic
interactions cause the rotational symmetry to be spontaneously broken without
long range positional or magnetic order. In fact, only one truly tetragonal
cuprate,(Ca,Na)$_{2}$CuO$_{2}$Cl$_{2}$, has been studied in this context, but
only by scanning tunneling microscope (STM) based spectroscopy, which is a
surface sensitive technique, and there is evidence that the surface may have a
lower symmetry due to a reconstruction (see below). Therefore the issue is the
more subtle one of whether there is an electronic instability of the CuO$_{2}$
planes towards a breaking of their four fold symmetry that drives the
pseudogap. This must be distinguished from the possibility that specific
probes are particularly sensitive to the pre-existing lack of four-fold
symmetry of the lattice. This is complicated by the fact that electronic
states around a gap or pseudogap are generally much more sensitive to
anisotropy than the more dispersive bands that would be present prior to the gapping.

\section{materials}

When discussing electronic instabilities in complex materials, such as
cuprates, it is important to keep in mind that electrons are charged
particles. Therefore they are subject to the Coulomb interaction both with
other electrons and with the ions that make up the crystal lattice. The
effective strength of the interaction between quasiparticles at the crystal
lattice is affected by screening. In fact, it is generally the case that
different electronic states and different energies have different couplings to
the lattice, as seen for example in mode and momentum dependent
electron-phonon couplings. There is no a priori reason for assuming that such
couplings are negligible, especially on low energy scales such as those
associated with the pseudogap.

The shadow Fermi surfaces seen \cite{aebi} in angle resolved photoemission
spectroscopy (ARPES) measurements provide an instructive example. A low
intensity replica of the Fermi surface of superconducting Bi$_{2}$Sr$_{2}%
$CaCu$_{2}$O$_{8+\delta}$ was observed shifted from the main Fermi surface by
($\pi/a$,$\pi/a$), which is the antiferromagnetic ordering vector of the
undoped cuprates. This shadow was ascribed to scattering by antiferromagnetic
fluctuations, which would be a novel electronic effect presumably related to
strong correlations. This led to a large body of follow on experimental and
theoretical work aimed at elucidating the detailed underlying mechanism.
However, it was pointed out early on that the shadow bands were remarkably
sharp for a fluctuation induced feature and that the lattices of the
Bi-cuprates have structural distortions that include the cell doubling needed
to produce this shadow. \cite{singh-bi1,singh-bi2} More recent experiments
showed that this prosaic picture explains the shadow Fermi surface.
\cite{shad,shad1,shad2} Importantly, this distortion lowers the symmetry from
tetragonal to orthorhombic and as evidenced by the observation of the shadow
Fermi surface couples significantly to the electronic states involved in
superconductivity and the pseudogap. Another instructive example is provided
by the high sensitivity of superconductivity in the (La,Ba,Sr)$_{2}$CuO$_{4}$
system to the structural phase transitions in particular the switching between
low temperature orthorhombic (LTO) to the low temperature tetragonal (LTT)
distorted structures with composition. \cite{crawford}

YBa$_{2}$Cu$_{3}$O$_{7-\delta}$ has two Cu-O systems, CuO$_{2}$ planes and
CuO$_{1-\delta}$ linear chains. The apical O atoms of the plane Cu are also
nearest neighbors of the chain Cu in this structure. This leads to
hybridization between the bands at the Fermi surface that are derived from the
plane Cu and electronic states associated with the chains, including near
$\delta$=0, the chain Fermi surface. \cite{pickett} The fact that there is
coupling between these two electronic subsystems is seen from the relatively
low anisotropy of the resistivity, $\rho_{c}/\rho_{ab}$, i.e. the relatively
high $c$-axis conductivity as compared to other layered cuprates and also the
opening of a substantial presumably induced superconducting gap on the chain
derived Fermi surface. As mentioned, the CuO$_{1-\delta}$
layer consists of Cu-O-Cu
chains. The chain formation is associated with O structural ordering, which
takes place at high $T$ and leads to an increasing orthorhombicity as the O
increasingly order with decreasing $T$ and annealing.\cite{veal,liang}
This chain formation persists to low doping (high $\delta$) and in fact
at $\delta=$0.5 an ordered orthorhombic
structure with alternating fully oxygenated chains and empty ``chains" forms.
High
resolution x-ray scattering studies have shown that there is an interplay
between this orthorhombicity and superconductivity, as reflected in an anomaly
in $b$/$a$ at $T_{c}$. \cite{horn} Importantly, substantial orthorhombicity
persists with increasing $\delta$ up to the superconductor--antiferromagnet
phase boundary. \cite{liang}

(Ca,Na)$_{2}$CuO$_{2}$Cl$_{2}$ is a true tetragonal material as was shown by
neutron diffraction measurements. \cite{argyriou} However, the only
experimental study pointing at nematicity in this material is by STM, which is
surface sensitive. A low energy electron diffraction (LEED) image of the
surface of this material was reported by Kohsaka and co-workers.
\cite{kohsaka} They do not find evidence for a reconstruction in their
pattern. However, based on the intensity of the Bragg peaks in the LEED
pattern, it was not of sufficient quality to detect a surface reconstruction
consisting of displacements of the surface atoms (see Refs.
\onlinecite{vanhove} and \onlinecite{heinz} for reviews of the LEED
technique). Moreover, a strong shadow band is seen in ARPES, and as in the Bi
compound discussed above, it is sharp in momentum space. This sharpness shows
that it comes from a static order rather than fluctuations, which would
introduce a width related to the momentum distribution of the fluctuations
that provide the scattering. This shadow is then a strong indication that the
surface is reconstructed. In fact, the ionic charges of Cl, Ca and Na are
almost certainly -1, +2 and +1, respectively. There is then no natural neutral
(001) cleavage plane in this crystal and so a reconstruction is highly likely.
More careful investigation of this surface is suggested to determine the
actual reconstruction pattern and its symmetry.

Finally, it is important to note that (1) an orthorhombic point group does not
contain a four-fold axis and therefore regardless of sample orientation,
measurement direction or coordinate system one cannot have a breaking of
four-fold symmetry, since there is no four-fold axis to start with; and (2)
the symmetry of a surface can be lower than the symmetry of the bulk, but it
cannot be higher; in particular a (001) face of an orthorhombic crystal will
not have four-fold symmetry.

\section{Scanning Tunneling Microscopy Based Spectroscopy}

We start with STM, as these experiments have been most emphasized in
connection with possible nematicity. The systems studied include Bi-2212
phases and (Ca,Na)$_{2}$CuO$_{2}$Cl$_{2}$. \cite{kohsaka-2,kohsaka-3,valla}
Importantly, (Ca,Na)$_{2}$CuO$_{2}$Cl$_{2}$ is the only cuprate that has been
studied in the context of nematicity at $T^{*}$ and is at the same time
crystallographically tetragonal in bulk. The measurements are done at low $T$,
well below $T^{*}$. Spectra are taken as a function of tip position and two
energy scales are identified in the spectra: $\Delta_{0}$, which is related to
superconductivity and $\Delta_{1}$, which is associated with the pseudogap.
$\Delta_{1}({\bf r})$ is found to vary considerably with position \textbf{r} on
the sample surface. This indicates a sensitivity of electronic states
$\Delta_{1}$ at the pseudogap edges to the local chemistry or structure
variations on the surface. This is as expected because states at and near a
gap edge are in general heavier than those associated with dispersive states,
and therefore more easily localized. This helps to confirm the association of
$\Delta_{1}$ with the pseudogap. Real space images of the spectra with the
intensity of $Z(\mathbf{r},E)$, which is the ratio of differential
conductances and opposite bias, defined in Ref. \onlinecite{kohsaka-3}, also
show variations, and these are correlated with the variations in $\Delta_{1}$,
so that real space plots of $f(\mathbf{r})=Z(\mathbf{r},\Delta_{1}%
(\mathbf{r}))$ show strong spatial dependence. What is being plotted then is a
function that has the most inhomogeneity in real space presumably because of
it has the most sensitivity to variations in local structure and chemistry.

As emphasized in the STM papers, this function shows chain-like correlations
running along the directions of the Cu-O-Cu bonds with lengths of several
atoms. As mentioned the structures of Bi-cuprates have strong lattice
distortions. They consist of shifts of O atoms in the Bi-O planes by $\sim$0.5
\AA , as well as large mainly transverse shifts of the apical O and buckling
of the CuO$_{2}$ planes, corresponding to tilts of the CuO$_{6}$ octahedra.
Such tilt type instabilities come from shifts of the O atoms and not the
cations. There is also an incommensurate modulation associated with the
mismatch between the Bi-O and CuO$_{2}$ systems. \cite{lepage} Further, the
Bi$^{3+}$-O$^{2-}$ surface is not charge neutral and therefore additional
distortions may be expected, although these have not been quantified
experimentally. In any case, tilt instabilities in perovskites are well
studied. \cite{glazer} Layered perovskite zone corner lattice instabilities
with tilts around [110] are close in energy to structures with [100] tilts.
Also, because of the bonding topology, in particular the square network of
corner sharing octahedra the tilts of nearby octahedra are generally
correlated. In other words, clockwise rotation of an octahedra drives
counterclockwise rotation of its neighbors because of the shared O atoms. This
is seen in the phonon dispersions of perovskites, where for example in cubic
perovskites the $R_{25}$ branch disperses weakly towards the $M$ point but
strongly towards $\Gamma$ (see e.g. Ref. \onlinecite{ghita} for a discussion
of this).

\begin{figure}[ptb]
\includegraphics[width=\columnwidth]{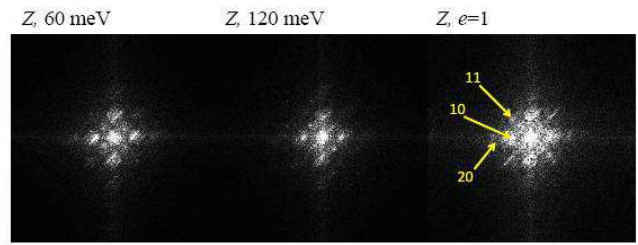}\caption{(color online)
FFT of Figs. 4a (left), 4b (middle) and 4e (right), from Ref.
\onlinecite{kohsaka-3}, with low order Bragg positions indicated. The FFT was
done on the figure converted to greyscale after trimming a border at the top
to remove the caption and slivers from the left and right to compensate the
top border thus yielding square images.}%
\label{fft-surface}%
\end{figure}

These tilts, which are strong in Bi-cuprates, provide a natural explanation
for chain-like correlations seen in the STM experiments if (1) local disorder
on the surface is strong enough to rotate the tilt axes away from [110]
towards [100] or [010] and (2) this couples to the electronic structure being
measured. In regard to (1), we note that the STM images for the underdoped
sample show substantial disorder, besides the chain-like correlations,
Visually, this appears stronger in the spectroscopic maps that show the
electronic structure of the Cu-O system, and is therefore sensitive to O
position than in the topographic image (Fig. 4d of Ref.
\onlinecite{kohsaka-3}, which shows the Bi positions) To quantify this, we
took the STM images of the spectroscopy $Z$ functions showing the chains
(Figs. 4a, 4b and 4e) of Kohsaka and co-workers, \cite{kohsaka-3} and Fourier
transformed (FT) it. The resulting transforms is shown in Fig.
\ref{fft-surface}. Only the first three Bragg reflections (100), (110), and
(200) are clearly seen and these are broad and fall off rapidly with order.
Note that even with strong variations in the intensities of the STM image from
site to site, if the atoms were on periodic sites Bragg peaks should be seen
up to higher order although they may be broadened, similar to what is seen in
x-ray crystallography for alloys. The FT instead indicates shifts of the atoms
from the periodically arranged sites in the bulk. In regard to (2) we note
that the Fermi surface comes from Cu $d_{x^{2}-y^{2}}$ - O $p_\sigma$
anti-bonding states, and the O $p$ orbitals ($x$,$y$,$z$) are orthogonal. This
is manifested in the shape of the Fermi surface which is flat near the nodal
directions. A square Fermi surface can be regarded as the intersection of
segments of two one dimensional Fermi surfaces, and so if one concentrates on
the nodal regions that provide the gap edge states around $\Delta_{1}$, one
has the intersection of two one dimensional surfaces, one along $x$ and the
other along $y$. The $\sigma$ hopping matrix elements are reduced
quadratically as the bond angle deviates from 180$^{\circ}$, providing a
mechanism for coupling of the gap edge states to the tilts. As noted, the
states at the Fermi surface are known to couple to the tilts as evidenced by
the shadow Fermi surface, and focusing on the states at the edge of the
pseudogap is expected to enhance this sensitivity as discussed. Therefore, the
images shown in the STM experiments can have a rather prosaic explanation.

Returning to chain-like correlations in perovskites, we mention that an
analogous situation can occur in the lattice dynamics of the soft mode of
ferroelectric perovskites such as BaTiO$_{3}$ and KNbO$_{3}$.
\cite{comes,krakauer} The real space interactions for the soft mode have a
one-dimensional character, leading to chain-like correlations seen in diffuse
scattering experiments above the Curie temperature. These correlations are of
comparable length to the chains in the STM images, although unlike them they
are dynamic rather than static. These are cubic crystals, not nematics, since
the correlated chains run equally along all three Cartesian directions.
Similarly, even though each molecule in a liquid of rod-like molecules may
have a certain orientation that correlates with its neighbors it is not a
liquid crystal unless the global rotational symmetry is lost.

Examining the FT image (Fig. \ref{fft-surface}), one notes that there is no
clear breaking of x - y symmetry. This means that the surface is not obviously
nematic on the scale of the STM image shown ($\sim$30 Cu spacings on a side).
One may ask if the image is a superposition of small nematic domains. In fact,
one can find regions in the image where two or more neighboring chains appear
to run together along the same direction. However, one can also find regions
with a single chain. The FT image shows diffuse intensity along the $H$0 and
0$K$ directions reflective of chain-like correlations (see Ref.
\onlinecite{comes}, but it is not clear that other streaks expected are
present, nor is there an strong asymmetry above the noise level in the shape
of e.g. the 10 Bragg peak that would indicate local nematic domains. We
conclude that at least in this image such regions, if present above a
statistical level, are very small.

\begin{figure}[ptb]
\includegraphics[width=\columnwidth]{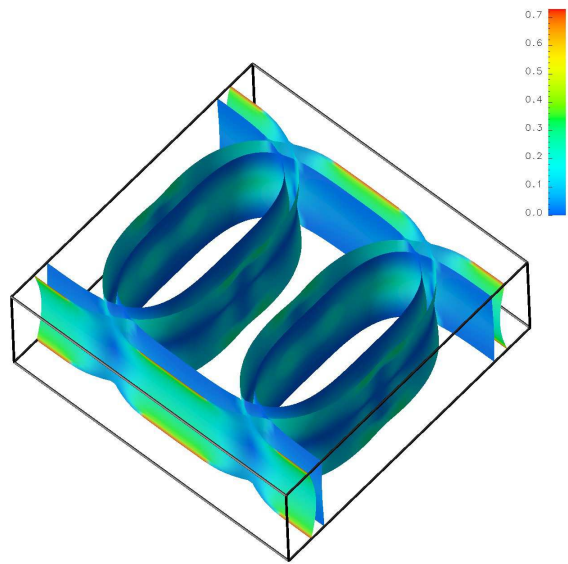}\caption{(color online)
Calculated Fermi surface of YBa$_2$Cu$_3$O$_{6.5}$ showing the
plane derived sections only (see text). There are two barrel shaped
sections corresponding to the two CuO$_2$ planes in the structure
and additionally a zone folding due to the alternation of filled and
empty chains in the CuO$_{1-\delta}$, $\delta=0.5$ layer.}
\label{ybco}%
\end{figure}


\section{Transport}

The Nernst coefficient, $\nu$, is a very sensitive probe of gapping.
\cite{behnia,chang} Like the thermopower, it is proportional to $T$ at low $T$
and in a simple band structure it is inversely proportional to the distance of
the band edge from the Fermi energy. Unlike the thermopower, $\nu$ is
proportional to the mobility, and electron and hole contributions are
additive. Large Nernst effects are found in semimetals, with Bi being a good
example, \cite{behnia} while, in normal metals the Nernst coefficient is very
small. Therefore, in a pseudogapped system one expects the Nernst coefficient
to grow strongly as regions of the Fermi surface in the nodal directions
are increasingly removed from participation in transport, consistent
with what was found experimentally. \cite{daou} On the level of the standard
Boltzmann kinetic theory with a single relaxation rate, the Nernst coefficient
is defined by derivatives of the transport function with respect to the
chemical potential, similarly to the Seebeck coefficient. \cite{behnia,ziman}
Thus, the Nernst coefficient is
particularly sensitive to those parts of the Fermi surface that most rapidly
change with the energy.
We show the plane derived sections of the Fermi surface of
YBa$_{2}$Cu$_{3}$O$_{6.5}$
in Fig. \ref{ybco}.
These were obtained with the general potential linearized augmented
planewave method \cite{singh-book}
as implemented in the WIEN2k code, \cite{wien}
using the
generalized gradient approximation of Perdew, Burke and Ernzerhof. \cite{pbe}
Our Fermi surface is similar to that reported previously by Carrington
and Yelland. \cite{carrington}
For the present discussion,
we have removed the chain derived Fermi surface from plot and
and colored the remaining Fermi surfaces according to the local
transport function, that is, the squared Fermi velocity.
We see that despite
the fact that the Fermi surface geometry retains the approximate tetragonal
symmetry, the transport function appears quite anisotropic, with some area of
heavy electrons at the \textquotedblleft y\textquotedblright\ faces, but not
\textquotedblleft x\textquotedblright\ faces. From this plot it is clear
that hybridization with chain Cu-O system
can make transport properties, especially the
Nernst coefficient, quite anisotropic, even if one excludes the transport in
the CuO chains themselves.

Indeed, Ando and co-workers \cite{ando} have measured the resistivity of
de-twinned YBa$_{2}$Cu$_{3}$O$_{7-\delta}$ crystals as a function of oxygen
content
and temperature. They find that the resistivity is anisotropic in all samples,
including $\delta$=0, and that in underdoped samples ($\delta$=0.55 and
$\delta$=0.35) the resistivity anisotropy increases with decreasing
temperature. The resistivity for high doping ($\delta$=0 and $\delta$=0.17) is
qualitatively different from the underdoped samples, presumably reflecting the
fact that the CuO$_{1-\delta}$ chains no longer contribute significantly to
the conductivity so that, as discussed by Ando and co-workers, the anisotropy
reflects anisotropy of the CuO$_{2}$ planes. Importantly, the anisotropy is a
smooth function of temperature up to the highest reported $T$=300K, which is
well above $T^{\ast}$, and furthermore shows no clearly
apparent structure that can be
associated with the pseudogap, although the pseudogap $T^{\ast}$ can be mapped
from analysis of the $T$ dependent curvature of the resistivity. \cite{ando2}
This is consistent with the view that the observed anisotropy is a
one-electron effect reflecting the anisotropy of the in-plane Fermi velocity
induced by the hybridization with chains.

On the other hand, the scattering rate may also be anisotropic (even if the
chain electrons are localized, they can still affect the scattering rates). In
fact, optical conductivity measurements suggest an anisotropy in the
scattering rate that behaves similarly above and below $T^{\ast}$. \cite{lee}
Either way, it is clear that the observed anisotropy in $\nu$ likely arises
from anisotropy in the electronic structure that exists independent of
$T^{\ast}$, which manifests itself particularly strongly in $\nu,$ at low $T,$
where the gapping is strong.
Therefore, while the Nerst anisotropy, may, of course, indicate a symmetry
breaking, a simpler explanation in terms of the always present
anisotropy of the underlying electronic structure, enhanced at low temperature
by gapping is not at all excluded.

\section{Neutron Scattering}

Inelastic neutron scattering offers a number of advantages. It is a true bulk
sensitive probe and it directly measures the dissipative part of the magnetic
susceptibility, $\chi^{\prime\prime}$ as a function of momentum and energy.
Hinkov and co-workers \cite{hinkov} reported such measurements for de-twinned
crystals of underdoped YBa$_{2}$Cu$_{3}$O$_{6.45}$. They showed by analysis of
the H,K,L dependent structure factors that the magnetic response that they
measure is from the CuO$_{2}$ planes and not the chain Cu. As expected they
find inelastic scattering near the (1/2,1/2) position corresponding to the
antiferromagnetic ordering vector of the undoped compounds. At high energy (50
meV) the distribution of $\chi^{\prime\prime} $ around this momentum is
isotropic but as the energy is lowered below 7 meV they find an anisotropy
that increases as the energy is lowered towards the quasielastic regime (3
meV). The scattering is elongated along the $a^{*}$ direction, indicating the
presence of chain-like magnetic correlations along the $b$ direction. This
anisotropy was found to be strongly $T$ dependent and vanishes below 100 K.
This very interesting behavior, i.e. the development of 1D magnetic
correlations in the CuO$_{2}$ planes as $T$ is lowered could be an indication
of nematicity, or it could be interpreted as nearness to magnetism in an
anisotropic (orthorhombic) material. Clearly, further investigation is needed.
However, one issue that should be emphasized is that the anisotropy is already
absent when the
temperature is raised to 100 K,
while $T^{*}$ for this composition is at much higher $T\sim
$250 K. \cite{ando2}
This calls into question the relationship between this anisotropy
and the pseudogap.
One experiment that would be particularly interesting would be inelastic
neutron measurements for a truly tetragonal material such as
(Ca,Na)$_{2}$CuO$_{2}$Cl$_{2}$ when suitable samples become available.

\section{summary}

In summary, we argue based on symmetry and other considerations that the
existing experimental data are not sufficient to prove a nematic origin of the
pseudogap in cuprates. Bulk sensitive measurements on tetragonal materials,
while difficult, will be very helpful in establishing whether the phase below
$T^{*}$ is associated with the onset of nematic order.

\acknowledgements

This work was supported by the Department of Energy, Materials Sciences and
Engineering Division (DJS) and the Office of Naval Research (IIM).

\bibliography{stm-3}

\end{document}